\documentclass[journal]{IEEEtran}

\usepackage{xcolor,soul,framed} 

\usepackage[pdftex]{graphicx}
\graphicspath{{../pdf/}{../jpeg/}}
\DeclareGraphicsExtensions{.pdf,.jpeg,.png}

\usepackage[cmex10]{amsmath}
\usepackage{booktabs}
\usepackage{hyperref}
\usepackage{rotating}
\usepackage{longtable,lscape}
\usepackage{algorithmic}
\usepackage{graphicx}
\usepackage{siunitx}
\usepackage{graphicx}
\usepackage{longtable,lscape}
\usepackage{multirow}
\usepackage{multicol}
\usepackage{amssymb} 
\usepackage{amsmath}
\usepackage{mathtools}
\usepackage{cite}
\usepackage{siunitx}
\usepackage{cuted}
\usepackage{flushend}
\usepackage{multibib}
\usepackage{relsize}
\usepackage{stix}
\usepackage{balance}
\usepackage{orcidlink}

\usepackage[utf8]{inputenc}
\usepackage[english]{babel}
\usepackage{array, makecell}

\DeclarePairedDelimiterXPP\BigOSI[2]{\mathcal{O}}{(}{)}{}{\SI{#1}{#2}}

\bstctlcite{IEEE:BSTcontrol}

\begin{document}
\bstctlcite{IEEEexample:BSTcontrol}
    \title{Accurate Loss Prediction of Realistic Hollow-core Anti-resonant Fibers Using Machine Learning}
  \author{ Yordanos Jewani,
           Michael Petry \orcidlink{0000-0002-8041-6246},~\IEEEmembership{Graduate Student Member,~IEEE,} ~\IEEEmembership{Student Member, Optica}, \\ Rei Sanchez-Arias\orcidlink{0000-0002-6145-7122}~\IEEEmembership{Member,~IEEE}, Rodrigo Amezcua-Correa,~\IEEEmembership{ Member,~IEEE,} ~\IEEEmembership{Member, Optica}, Md Selim Habib \orcidlink{0000-0003-0161-5325},~\IEEEmembership{Senior Member,~IEEE,} ~\IEEEmembership{Senior Member, Optica}

  \thanks{Manuscript received August $\times\times$, 2023; $\times\times$, 2023; accepted $\times\times$, 2023. Date of publication $\times\times$,
2023; date of current version $\times\times$, 2023. 
This work was supported in part by the Woodrow W. Everett, Jr. SCEEE Development Fund in cooperation with the Southeastern Association of Electrical Engineering Department Heads and the U.S. Department of Treasury under the Coronavirus State and Local Fiscal Recovery Funds. M. Petry acknowledges support from U.S. Fulbright Scholarship.
(\emph{Corresponding author:
Md Selim Habib.})}
   \thanks{
   Y. Jewani and R. Sanchez-Arias are with the Department of Data Science and Business Analytics, Florida Polytechnic University, 4700 Research Way, Lakeland, FL-33805, USA (e-mail: yjewani9995@floridapoly.edu and  rsanchezarias@floridapoly.edu).
   
   M. Petry is with the Faculty of Electrical Engineering and Information Technology, University of Applied Sciences Karlsruhe, BW-76133, Germany (e-mail: michael.petry@fulbrightmail.org).
   
   R. Amezcua-Correa is with the
CREOL, University of Central Florida, Orlando, FL 32816 USA (e-mail: r.amezcua@creol.ucf.edu).

   M. Selim Habib is with the Department of Electrical Engineering and Computer Science, Florida Institute of Technology, Melbourne, FL-32901, USA (e-mail: mhabib@fit.edu).
   
   }
   }

\markboth{IEEE Journal of Selected Topics in Quantum Electronics, VOL.~XXX, NO.~XXX, August~2023}
{Petry \MakeLowercase{\textit{et al.}}:XXX}

\maketitle


\begin{abstract}
Hollow-core anti-resonant fibers (HC-ARFs) have proven to be an indispensable platform for various emerging applications due to their unique and extraordinary optical properties. 
However, accurately estimating the propagation loss of nested HC-ARFs remains a challenging task due to their complex structure and the lack of precise analytical and theoretical models. To address this challenge, we propose a supervised machine-learning framework that presents an effective solution to accurately predict the propagation loss of a 5-tube nested HC-ARF. Multiple supervised learning models, including random forest, logistic regression, quadratic discriminant analysis, tree-based methods, extreme gradient boosting, and K-nearest neighbors are implemented and compared using a simulated dataset. Among these methods, the random forest algorithm is identified as the most effective, delivering accurate predictions. Notably, this study considers the impact of random structural perturbations on fiber geometry, encompassing random variations in tube wall thicknesses and tube gap separations. In particular, these perturbations involve randomly varying outer and nested tube wall thicknesses, tube angle offsets, and randomly distributed non-circular, anisotropic shapes within the cladding structure.  It is worth noting that these specific perturbations have not been previously investigated. Each tube exhibits its unique set of random values, leading to longer simulation times for combinations of these values compared to regular random variables in HC-ARFs with similar tube characteristics. The comprehensive consideration of these factors allows for precise predictions, significantly contributing to the advancement of HC-ARFs for many emerging applications.

\end{abstract}


\IEEEpeerreviewmaketitle

\section{Introduction}

\IEEEPARstart{T}{he} advent of optical fiber technology has sparked a revolutionary transformation in long-distance data transmission, profoundly impacting various industries \cite{kaushik2020fibre}. By utilizing high-purity glass that minimizes signal loss, this technology enables the transmission of data through light, achieving exceptionally high-speed communication over long distances \cite{tamura2018first}. A notable advancement in this field is the utilization of hollow-core anti-resonant fibers (HC-ARFs), which employ a hollow-core surrounded by closely spaced anti-resonant tubes to guide light in the air-core with low loss, low power overlap with silica parts, and wider transmission bandwidth \cite{pryamikov2011demonstration,yu2012low,kolyadin2013light,yu2013spectral,belardi2014hollow,poletti2014nested,habib2015low,uebel2016broadband,wei2017negative,frosz2017analytical,debord2017ultralow,gao2018hollow,habib2019single,jasion20220,habib2021impact,amrani2021low, wang2023multi,osorio2023hollow}. In contrast to solid-core fibers, HC-ARFs utilize an unconventional and remarkable guiding mechanism called inhibited-coupling between the core modes and cladding modes, along with the anti-resonant effect, to facilitate strong and well-controlled light propagation within the air-core \cite{couny2007generation}. The exceptional optical characteristics of HC-ARFs give rise to a wide range of desirable applications. These include short-reach data transmission \cite{sakr2020interband,sakr2019ultrawide}, quantum state transmission \cite{ding2019recent}, polarization purity \cite{Taranta2020Exceptional}, polarization control \cite{habib2021enhanced,hong2022highly}, high-power delivery \cite{michieletto2016hollow,gebhardt2017nonlinear,mulvad2022kilowatt,cooper2023}, extreme nonlinear optics \cite{travers2011ultrafast,russell2014hollow,adamu2019deep,adamu2020noise,habib2017soliton,markos2017hybrid,cassataro2017generation,sollapur2017resonance}, optofluidic \cite{hao2018optimized}, terahertz transmission \cite{anthony2011thz,hasanuzzaman2015low,sultanan2020exploring}, low-noise applications \cite{iyer2020ultra}, and to mention a few. 
One of the primary performance factors is the accurate calculation of propagation loss in HC-ARFs. Several approaches have been utilized for calculating the loss of HC-ARFs, including analytical models \cite{ding2014analytic, vincetti2016empirical}, semi-analytical models \cite{ding2015semi}, and numerical methods such as the finite-element method (FEM) \cite{cucinotta2002holey}. While the analytical and semi-analytical models are suitable for simplified fiber structures such as single-ring HC-ARFs \cite{zeisberger2017analytic} and stadium-shape HC-ARFs \cite{murphy2023azimuthal}, achieving accurate modeling of HC-ARFs usually requires the extensive and time consuming FEM modeling. However, precisely estimating the propagation loss in realistic nested HC-ARFs remains challenging due to their intricate structure and the lack of precise analytical and theoretical models. Taking the HC-ARF platform to the next step requires precise loss prediction with feasible computational time.

Despite of numerous opportunities and ubiquitous applications, there has been a limited number of research investigations focusing on the utilization of machine learning methods to predict the loss of HC-ARFs \cite{hu2020design, meng2021use}. 
\begin{figure*}[t!]
  \begin{center}
  \includegraphics[width=6.2in]{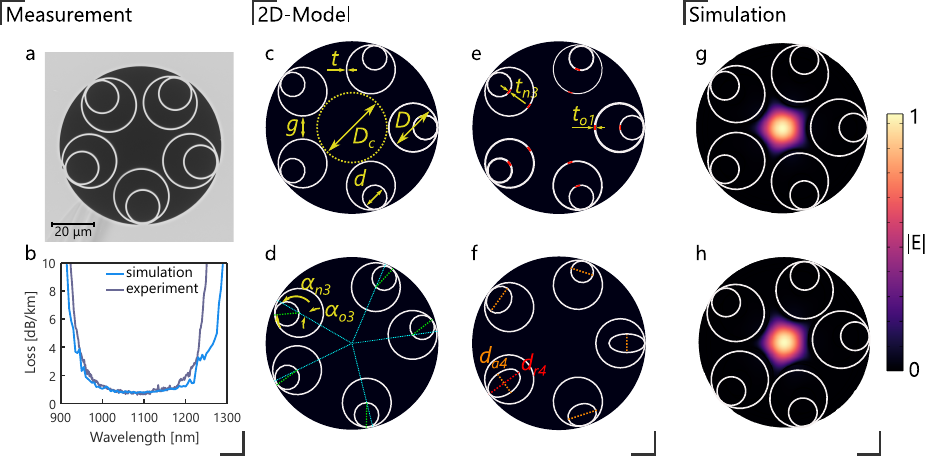}

  \caption{Measurement: (a) scanning electron microscrope (SEM) image of a fabricated 5-tube nested HC-ARF and  (b) Simulated (cyan) and measured (gray) loss spectrum. The fiber was fabricated in-house at CREOL, University of Central Florida, USA \cite{cooper2023}. The fiber has a core diameter of $\approx$23 $\mu$m and average wall thickness of outer tubes and nested tubes of 780 nm $\pm$ 10 nm.   2D-Model (c-f): cross-sections of ideal and perturbed 5-tube nested HC-ARF geometries. Variables correspond to design parameters in Table \ref{tab:fibergeometrics}. Random perturbations include (d) angular offset of nested and outer cladding elements, (e) random silica wall thickness of nested and outer cladding elements, and (f) anisotropic stretched nested cladding elements in radial and axial direction. Illustrations exaggerate perturbation intensity for enhanced visibility. Simulation (g-h): mode-field profiles of (g) ideal fiber geometry, and (h) exemplary perturbed fiber geometry.}
  \label{fig:crossections}
  \end{center}
\end{figure*}
For example, a reinforcement learning methodology was employed to effectively optimize the design parameters of HC-ARF  \cite{hu2020design}. More recently, machine learning models were also employed to anticipate the confinement loss of HC-ARF \cite{meng2021use}, in which the study utilizes the structure-parameter vector of ARFs with single and double layers of elliptical cladding tubes, where the ARF structures are defined by various structural styles of $s_1$ and $s_2$, with a range of 5-10 first/second tubes ($N$). The magnitude of confinement loss was predicted using classification algorithms such as decision trees and K-nearest neighbors. However, previous machine learning related studies did not consider the impact of random structural perturbations on the fiber geometry that will inevitably be imprinted on the fiber during the manufacturing process. As first reported in \cite{petry2022random}, these include random variations in silica tube wall thicknesses, tube gap separations, angular misalignments of both outer and inner cladding elements, and anisotropic deformation effects. Since each fiber sample exhibits its own, individual combination of random perturbations, separate finite-element analyses have to be performed to calculate propagation characteristics individually, which poses a drastically time-consuming task.

This work aims to accurately and efficiently predict the propagation loss of randomly structured nested HC-ARFs using supervised machine learning algorithms, and to assess and compare the performance of various algorithms. Given that classification algorithms offer techniques to handle imbalanced data, where one class may have significantly more samples than others, we apply the synthetic minority oversampling technique (SMOTE) as a resampling method to balance the distribution of the simulation data used for training the machine learning models. 

The article is organized as follows: Section \ref{sec:hcarf_architecture} provides an overview of the architecture of HC-ARF geometry and the simulation environment. Section \ref{sec:methodology} explores the application of machine learning models in fiber optics, discussing various supervised machine learning algorithms such as random forest, logistic regression, quadratic discriminant analysis, tree-based methods, extreme gradient boosting, and K-nearest neighbors, as well as the use of SMOTE to address the issue of imbalanced data. Section \ref{sec:results} presents the results of the machine learning models and compares their performance. Finally, Section \ref{sec:conclusion} summarizes the key points discussed in this article and provides future directions.

\section{HC-ARF Architecture}
\label{sec:hcarf_architecture}
This section provides a detailed look into the HC-ARF architecture under investigation, discusses corresponding random geometric anomalies observed during the manufacturing process, and introduces FEM simulation environment used to calculate ground truth propagation characteristics throughout this work. 

\subsection{Fiber Geometry}

In this study, all geometric characteristics described are defined based on the cross-section of the HC-ARF, as depicted in Fig. \ref{fig:crossections}, which is divided in three sections: \textit{Measurement} (a-b), \textit{2D-Model} (c-f), and \textit{Simulation} (g-h). Scanning electron microscope (SEM)  image of a fabricated 5-tube nested HC-ARF similar to the one studied here are provided in Fig. \ref{fig:crossections}(a) with the purpose of motivating this work. The fiber has a core diameter of $\approx$23 $\mu$m, while the average wall thickness of both the outer and nested tubes is 780 nm with a variation of $\pm$10 nm. It can be seen from the SEM image that the silica cladding tubes exhibit imperfections and deformations. Such fiber imperfections are also reported in the previous studies \cite{sakr2019ultrawide,sakr2020interband}.  The most distinctive anomaly is a misalignment of the nested tubes with respect to their corresponding outer tube, resulting in a rolling effect as first described in \cite{petry2022sum}. 
In a measurement reported in \cite{PetryOE22}, the magnitude of random deviations observed in realistic fibers has been analyzed on a per-lot basis for certain labeled geometric properties. This analysis has provided a range of realistic magnitudes for each variable. Table \ref{tab:fibergeometrics} summarizes these ranges together with the selected base geometric properties such as the wavelength $\lambda$, core diameter $D_\text{c}$ and outer tube gap separation $g$. In 2019, Habib \textit{et al.}  \cite{habib2019single}  showed that the number of cladding tubes plays a crucial role in designing wide-band and ultra-low loss fibers with effectively single-mode operation. The 5-tube HC-ARF was proven to excel in certain propagation characteristics, including but not limited to a wide transmission window and a remarkably low loss compared to other fiber geometries, which can be seen from the measured loss values in Fig. \ref{fig:crossections}(b) of the fabricated fiber. The loss measurement was performed by coupling a white light source (NKT SuperK COMPACT) into a 463 m long fiber and then cut to a 100 m on a 30 cm diameter coil, while maintaining constant input conditions. The loss measurement shows that the fiber has a wider transmission bandwidth and record low loss of 0.79 dB/km at 1080 nm. It can be seen from Fig. \ref{fig:crossections}(b) that the simulated loss agrees well with the measured loss values. The simulations were performed using the approach depicted in \cite{habib2019single}. In this study, a 5-tube nested HC-ARF geometry is chosen due to the outstanding optical properties. One of the unique features of HC-ARFs is that the transmission window can be shifted by properly choosing the silica wall thickness. 

In total, six geometric anomalies are considered in this study, which are depicted in Fig. \ref{fig:crossections}(d-f). In this order, these are an individual random angular offset of both the outer and nested cladding elements, denoted by $\alpha_\text{0i}$ and $\alpha_\text{ni}$, respectively, an individual random outer and nested tube silica thickness, denoted by $t_\text{0i}$ and $t_\text{ni}$, respectively, and an individual stretch to the nested tubes in both radial and axial directions, denoted by $d_\text{ri}$ and $d_\text{ai}$, respectively. As indicated by the subscript $i$, which ranges from $1$ to $5$, each cladding element has its individual variable assigned. Prior to every simulation run, the magnitudes for each of these six effects are determined by sampling a standard deviation uniformly from the ranges given in rows five to ten in Table. \ref{tab:fibergeometrics}. Then, for each effect, five samples are drawn from a normal distribution using the prior sampled standard deviation, which are then applied on the five cladding elements. This procedure is repeated for every simulation to guarantee the generation and analysis of a new, unseen geometry, leading to a high diversity in the data set. In total 65928 simulation runs have been performed using extensive FEM simulations. To solve the eigen-mode problem of a realistic HC-ARF, a considerably large computational domain and substantial time are required. This is due to the utilization of the entire fiber domain for loss calculations. For example, each sample simulation took approximately three minutes on a desktop computer equipped with an Intel\textsuperscript \textregistered Xeon\textsuperscript \textregistered Platinum 8260 CPU @ 2.40GHz (2 processors) and 64.0 GB RAM.  Therefore, optimization is crucial to simultaneously reduce the computational time and accurately model the propagation loss and modal contents of HC-ARF.

\subsection{Simulation Environment}
The simulations were carried out using a commercially available FEM-based \textsc{Comsol}$^{\circledR}$ Multiphysics software combined with MATLAB-Livelink environment. The silica structure of the fiber has been modeled using triangular mesh-elements with maximum edge-lengths of $\lambda/6$, whereas maximum edge-lengths of $\lambda/4$ are used for the air-regions similar to \cite{poletti2014nested,habib2019single,habib2021impact,jasion20220}. To approximate an infinitely extended air-region around the fiber, at least 10-layer-deep perfectly matched layer with optimized boundary conditions were used similar to \cite{habib2019single,habib2021impact}. This study focuses on the fundamental-mode propagation loss ($LP_{01}$-like mode) for both polarizations: $LP_{01}^x$ and $LP_{01}^y$. 
Since the geometric anomalies lead to an asymmetrical structure, solving for both $LP_\text{01}$-modes is necessary. To decrease mode-searching time further, the initial refractive mode index is approximated using an analytical capillary model \cite{marcatili1964hollow}: $n_\text{guess}=\sqrt{1-\left(\frac{U_{mn}\lambda}{2\pi R_c}\right)^2}$, where, $R_c$ is the core radius, $\lambda$ is the wavelength, and $U_{mn}$ is the $n$th zero of the $m$th-order Bessel function of the first kind.

To calculate the total fiber propagation loss, confinement or leakage losses (CL) as well as surface scattering losses (SSL) are considered. The SSL is estimated using \cite{poletti2014nested}: $\alpha_\text{SSL} [\si{dB/km}]=\eta{}F \left(\frac{\lambda}{\lambda_0}\right)^{-3}$  with the $F$-factor as described in \cite{fokoua2012scatter}. Furthermore, choosing $\eta=150$ calibrates the estimation for $\lambda_0=\SI{1.55}{\micro\meter}$ \cite{fokoua2023loss}. Considering the wavelength dependency of this loss is not required, since the fiber analysis happens at the fixed wavelength $\lambda_0$. Furthermore, effective material loss (EML) is neglected because of the insignificant power overlap with silica glass  $<10^{-4}$.

\section{Methodology}
\label{sec:methodology}

\begin{figure}[t!]
	\centering
	\includegraphics[width=3.2in]{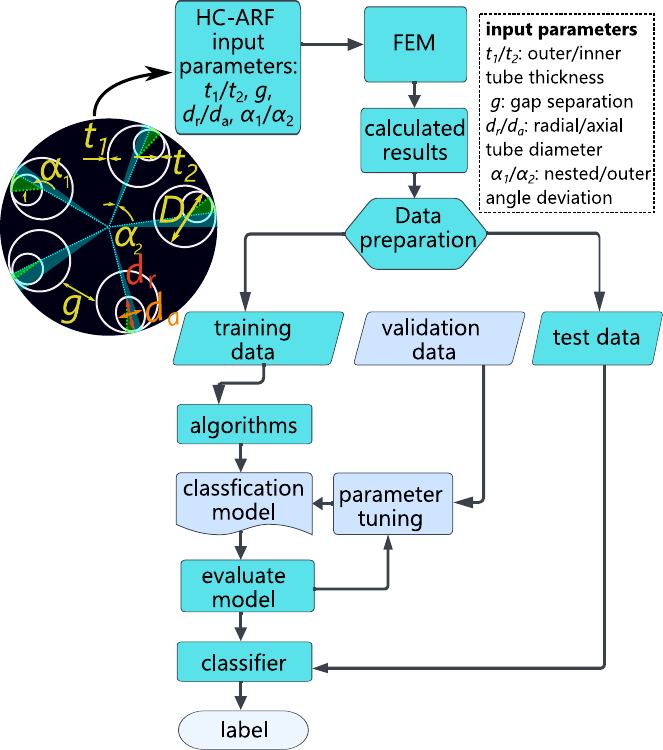}
	\caption[Machine learning-based flowchart for predicting the total loss of HC-ARF]{Machine learning-based flowchart for predicting the total loss of HC-ARF: (1) simulation data, (2) data pre-processing including normalizing and scaling data, encoding categorical variables, and data cleaning such as missing values, (3) data preparation: splitting data into a training set (training and validation data) and test set, and (4) model building and evaluation.}
	\label{fig:figure-1}
\end{figure}

In this study, supervised machine learning algorithms are used to predict the propagation loss of HC-ARF by employing FEM simulated data. The data generated from FEM simulations is fed as input to a data pre-processing stage and later split into training and testing sets. The training set again is split into training and validation sets. Several models, including random forest, logistic regression, quadratic discriminant analysis (QDA), tree-based methods, extreme gradient boosting (XGboost), and K-nearest neighbors (KNN) are considered within the scope of this work. Validation data is used to measure the trained models' accuracy and tune the parameters of the models as needed. Then the final model is used to classify the data into the different categories. The general flow of the above process is shown in Fig. \ref{fig:figure-1}.

\subsection{Machine Learning Models}
\label{subsec:machine_learning_techniques}

Supervised learning is a type of machine learning where an algorithm learns from labeled data to make predictions or classify new data. Binary classification is a common classification task that learns to classify input data into one of two classes. This section discusses various classification techniques, including tree-based methods, and describes how we deal with imbalanced data in classification tasks \cite{hastie2009elements}.

Logistic regression is an extensively used statistical learning method, especially for tasks where there are only two class values, that model the probability of a discrete outcome by fitting a logit function to the dataset \cite{bisong2019logistic}. The KNN method is a classification algorithm that aims to determine the class label of a test instance based on the majority class labels of its K-nearest neighbors in the training dataset \cite{zhang2017learning}. On the other hand, QDA method \cite{siqueira2017lda} is a type of generative statistical model that assumes each class has its own multivariate Gaussian distribution with distinct mean vectors and covariance matrices. QDA uses these Gaussian distributions to estimate the class conditional probabilities.

In decision tree methods the data is continuously split according to a specific parameter to generate a predicted label. The tree can be defined by two entities, namely decision nodes, and leaves. The leaves are the decisions or the final outcomes, and the decision nodes are where the data is split\cite{hastie2009elements}. Random forests (RF) is an ensemble learning method that combines the predictions of multiple decision trees to improve accuracy and robustness in classification tasks \cite{breiman2001random}. Leo Breiman explains \cite{breiman2001random} that RF creates an ensemble of decision trees by using a combination of bagging (Bootstrap Aggregating) and feature randomization. J. H. Friedman introduced the concept of the gradient boosting method (GBM), which is an iterative technique that combines multiple models to improve prediction accuracy \cite{friedman2001greedy}. Extreme gradient boosting (XGBoost) is an efficient implementation of the GBM algorithm \cite{chen2016xgboost}. It was developed by Tianqi Chen and Carlos Guestrin \cite{10.1145/2939672.2939785} and is now one of the most widely used machine learning algorithms. XGBoost uses a combination of gradient descent optimization and parallel processing to train models quickly and efficiently.

\begin{table}[t!]
\centering
\caption{Parameter description of 5-tube nested HC-ARF.}
\label{tab:fibergeometrics}
\resizebox{0.49\textwidth}{!}{%
\begin{tabular}{@{}llll@{}}
\toprule
No. & Symbol                & Range                 & Description                                    \\ \midrule
1   & $\lambda$                   & 1.55 $\mu$m               & wavelength                                     \\
2   & $D_\text{c}$                    & 35 $\mu$m           & core diameter                                    \\
3   & $N$                     & 5                     & number of tubes                                \\
4   & $g$                 & 5.25 $\mu$m               & gap separation                                 \\
5   & $t_0$      & U\{393 nm, 0--10\%\} & outer tube thickness                    \\
6   & $t_n$      & U\{393 nm, 0--10\%\} & nested tube thickness                  \\
7   & $\alpha_0$  & U\{0--2°\}           & outer tube angle offset                 \\
8   & $\alpha_{n}$ & U\{0--15°\}          & nested tube  angle offset                \\
9   & $d_a$      & U\{0--20\%\}                & anisotropical axial tube  \\
10  & $d_r$      & U\{0--20\%\}                & anisotropical radial tube  \\ \bottomrule
\end{tabular}%
}
 \label{tab:table1}
\end{table}


The dataset exhibits significant class imbalance with the majority class (high loss) accounting for 90\(\%\) of the observations and the minority class (low loss) representing only 10\(\%\). This imbalance may negatively impact the performance of machine learning models.  To address this issue, we use a data augmentation technique called SMOTE, a methodology widely adopted in machine learning for dealing with imbalanced data. It generates synthetic minority class examples by interpolating between existing minority class examples. Specifically, SMOTE randomly selects a minority class example, identifies its K-nearest neighbors, generates artificial examples by interpolating between the selected sample and its neighbors, and finally under-samples the majority class. This way, the SMOTE algorithm increases the size of the minority class and helps to balance the class distribution. This improves the performance of machine learning models that are trained on imbalanced datasets \cite{chawla2002smote}.

\subsection{Dataset and Models}
\label{subsec:data_r}
\begin{table}[t]
\centering
\caption{Boundary of the Binary propagation loss class.}
\label{tab:table2}
\resizebox{0.42\textwidth}{!}{
\begin{tabular}{@{}lll@{}}
\toprule
\multicolumn{2}{c}{Classification Boundary}        & \multicolumn{1}{c}{Class} \\ \midrule
Binary-class & total loss $\textless$ \SI {0.15} {dB/km} & low loss                \\
             & total loss  $\geq$ \SI {0.15} {dB/km}     & high loss                 \\ \bottomrule
\end{tabular}
}
\end{table}


In this study, randomly assigned values of fiber structure parameters as displayed in Table \ref{tab:table1} are used as an input to a FEM simulation. The propagation loss (CL and SSL), effective mode index ($n_\text{eff}$), power overlap: $\eta = \frac{P_\text{core}}{P_\text{all}}$, where $P_\text{core}$ is the power in the core, and  $P_\text{all}$ is the power in the whole structure, and the $F$-factor were calculated using FEM. The objective is to predict the magnitude of the total propagation loss and treat it as a classification task to categorize the total loss into different classes based on the fiber's structure, properties, and application. The dataset used in this study after the SMOTE technique has 80000 instances with seven features, including the response variable "total loss", and a binary classification threshold is applied as shown in Table \ref{tab:table2}.

The R programming language is used to implement all the machine learning methods studied in this work using \texttt{\{tidymodels\}}, an integrated framework that aims to provide a complete workflow of data preprocessing, feature engineering, model selection, and evaluation with a consistent approach. \texttt{\{tidymodels\}} is a comprehensive ecosystem of R packages designed for machine learning and modeling tasks, following the principles of the \texttt{\{tidyverse\}}, a modern and powerful collection of packages for data science. This allows for a consistent approach to building, training, evaluating, and interpreting multiple machine learning models \cite{kuhn2020tidymodels}.

The R package \texttt{\{resample\}} is used for handling data splitting and resampling tasks. It provides functions for creating training and test sets, as well as performing multiple types of resampling techniques such as cross-validation and bootstrapping. Using functions from this package  our simulation data is split into training and test sets using a 70\(\%\):30\(\%\) ratio and a standard 10-fold cross-validation is performed on the training set \cite{kuhn2020tidymodels,chihara2022mathematical}.

\begin{table*}[ht]
\centering
\caption{Comparison of Performance Metrics for Different Algorithms \cite{kuhn2013applied}.}
\label{tab:table4}
\resizebox{0.8\textwidth}{!}{%
\begin{tabular}{lcccccc}
\hline
\multicolumn{1}{c}{\textbf{Model}} & \textbf{Accuracy} & \textbf{ROC AUC} & \textbf{Sensitivity} & \textbf{Specificity} & \textbf{Precision} & \textbf{F1}    \\ \hline
Decision tree       & 0.663 & 0.718 & 0.769 & 0.521 & 0.682 & 0.723 \\
KNN                 & 0.847 & 0.929 & 0.826 & 0.874 & 0.897 & 0.86  \\
Logistic regression & 0.594 & 0.604 & 0.844 & 0.262 & 0.604 & 0.704 \\
QDA                 & 0.597 & 0.615 & 0.818 & 0.303 & 0.61  & 0.699 \\
XGBoost             & 0.790  & 0.862 & 0.854 & 0.704 & 0.794 & 0.823 \\
\textbf{Random forest}             & \textbf{0.868}    & \textbf{0.936}   & \textbf{0.901}       & \textbf{0.823}       & \textbf{0.871}     & \textbf{0.886}\\

\bottomrule
\end{tabular}%
}
\end{table*}

The supervised machine learning models studied in this work are built using the R package \texttt{\{parsnip\}} \cite{kuhn2020tidymodels}, useful for model specification and tuning. It provides a simple and intuitive syntax for model specification using a "model formula" approach using a unified syntax for all methods. Additionally, it supports model tuning, hyper-parameter optimization, and ensemble learning techniques, making it a comprehensive tool for building and evaluating machine learning models. The model evaluation and performance metrics are carried out using the \texttt{\{yardstick\}} package, which provides a  collection of functions for calculating and visualizing evaluation metrics to assess the performance of machine learning models \cite{yardstick}. Accuracy, precision, recall, F1 score, the receiver operating characteristic (ROC) area under the curve (AUC), among other performance metrics for classification tasks are computed in this work.

\section{Results and Discussion}
\label{sec:results}

In this section, we provide an overview of the performance of the  machine learning models studied in this work, as well as a discussion comparing the promising supervised learning algorithms tested here. We present the findings from our evaluation and discuss the implications and insights gained from these results. A brief description of the metrics used to assess the accuracy of the applied machine learning algorithms is included next. 

\subsection{Performance Metrics}
\label{subsec:performance_measurement}

\noindent\textbf{Accuracy} is the correct number of predictions the model makes for all observed values. Accuracy is computed as shown in (\ref{eu_eqn1}), where TP refers to true positives, TN to true negatives, FP to false positives, and FN to false negatives \cite{9121169}:

\begin{equation} \label{eu_eqn1}
\text{Accuracy} = \frac{\text{TP + TN}}{\text{TP + TN + FP + FN }}. 
\end{equation}

\noindent\textbf{Precision} is a measure of the accuracy of a classifier in correctly categorizing instances of a specific class out of all the instances categorized as that class. It is calculated using the relation shown in (\ref{eu_eqn2}):

\begin{equation} \label{eu_eqn2}
\text{Precision} = \frac{\text{TP}}{\text{TP + FP}}.
\end{equation}         

\noindent\textbf{Recall}, also known as true positive rate or \textcolor{black}{\textbf{sensitivity}}, measures the proportion of true positive cases that are correctly identified by the test \cite{8895818}, that is:

\begin{equation} \label{eu_eqn3}
\text{Recall} = \frac{\text{TP}}{\text{TP + FN}}. 
\end{equation}  

\noindent\textbf{Specificity} measures the proportion of true negative cases correctly identified as negative by the test:

\begin{equation} \label{eu_eqn4}
\text{Specificity} = \frac{\text{TN}}{\text{TN + FP}}.
\end{equation} 

\noindent Since precision and recall individually do not cover all aspects of accuracy, we take their harmonic mean to compute the \textbf{F1-Score}, as shown in (\ref{eu_eqn5}), which covers both aspects and better reflects the overall measure of accuracy. It ranges from 0 to 1 and can be computed by:

\begin{equation} \label{eu_eqn5}
\text{F1-Score} = 2 \times \frac{\text{Precision} \times \text{Recall}}{\text{Precision + Recall}}. 
\end{equation}

\begin{figure}[b!]
	\centering
 \includegraphics[width=0.9\linewidth]{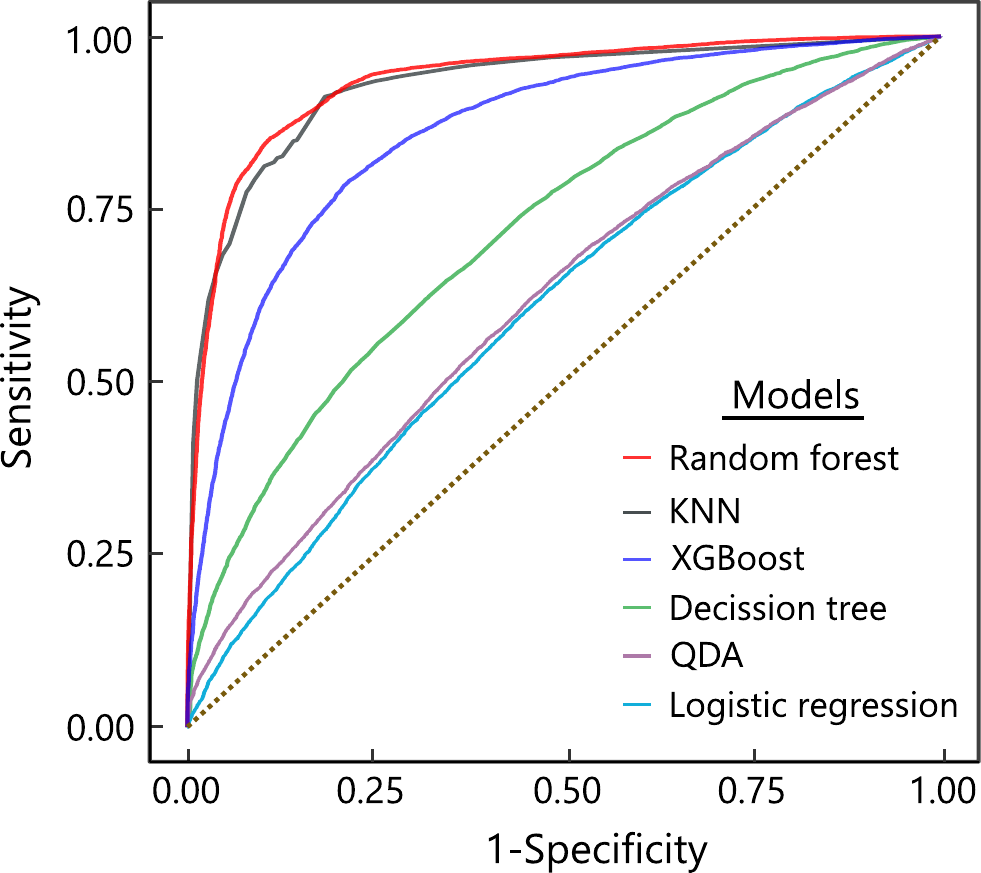}
	\caption[Machine learning model comparison using ROC curve analysis. Illustrating the ROC curve analysis for data with imbalanced and balanced distributions using the SMOTE (Synthetic Minority Over-sampling Technique) method.]{Machine learning models comparison using receiver operating characteristic (ROC) curve analysis. The relationship between the True Positive Rate (TPR), also known as Recall or Sensitivity, and the False Positive Rate (FPR), calculated as (1-Specificity) is generated for multiple thresholds. TPR is represented in the vertical direction and FPR in the horizontal direction. Diagonal line is included to represent random classification with equal chances of true positives and false positives. A curve close to the diagonal line,  suggests that the classifier performs poorly and is not better than random chance. An ROC curve closer to the top-left corner suggests a model with high predictive capabilities.} 
	\label{fig:figure-3}
\end{figure}

Our study utilizes classification algorithms to predict the propagation loss of five tube HC-ARF, with classification labels defined in Table \ref{tab:table2}. Among the multiple algorithms tested, random forest (RF) demonstrated superior performance, achieving an accuracy of 0.868, precision of 0.87, recall/sensitivity of 0.901, ROC-AUC score of 0.936, and F1 score of 0.886, as reported in Table \ref{tab:table4}. These results suggest that RF is the most suitable algorithm for the classification task and type of dataset of interest in this study. The random forest algorithm outperformed other methods in accurately identifying positive and negative instances.

\subsection{ROC Curve and Discussion}
\label{subsec:confusion_matrix_roc_curve}

A receiver operating characteristic curve (ROC curve) is commonly used to study the performance of a binary classification model at all classification thresholds. It shows the relation between true positive rate (Sensitivity) and false positive rate (1 - Specificity). Therefore, the ROC area under the curve (AUC), provides an aggregate measure of performance across all possible classification thresholds. The AUC can be interpreted as the probability that the model will rank a randomly chosen positive example more highly than a randomly chosen negative example. The ROC curve provides an effective visual comparison of the performance of different classifiers, and is shown in Fig. \ref{fig:figure-3}. The random forest model had a ROC AUC score of 0.93 and 86\(\%\) accuracy, and the performance of this model compared to other methods is evident as shown in  Fig. \ref{fig:figure-3}.


This finding suggests that random forest is a well-suited algorithm for predicting the propagation loss with the classification task and dataset. Computation of variable importance for the random forest algorithm \cite{kuhn2013applied}, indicated that the thickness of the outer tube was one of the most influential factors in building the decision trees, along with the anisotropic radial tube parameters. The remaining parameters had similar contribution in the generation of the decision trees for this ensemble model.

\section{Conclusion and Future Directions}
\label{sec:conclusion}

This study tackles the challenges associated with predicting propagation loss of HC-ARFs using traditional numerical methods by proposing a novel approach based on supervised machine learning. The proposed approach involves binary-classification tasks using data generated from FEM simulations of HC-ARFs. The machine learning models are trained on a labeled dataset of FEM simulation results, considering input parameters relevant to HC-ARF design, including the design structure of the HC-ARF used for this study. Varying outer and nested tube wall thicknesses, tube angle offsets, as well as anisotropic radial and axial properties are part of our design. Performance evaluation metrics, including accuracy, sensitivity, specificity, F1-score, and ROC curve analysis, are used to assess the prediction capabilities of the models. Random forest emerged as the most effective technique, demonstrating high accuracy in predicting the type of propagation loss. Future work could explore a multi-class classification setting and unsupervised learning models to gain further insights into HC-ARF performance that can help in understanding other complex attribute relationships within the model. \textcolor{black}{The work presented in this paper demonstrates the ability of machine learning (ML) methods to effectively classify a loss level. Our framework and results illustrate the supervised learning problem of classification and compare ML methods of different complexity used to predict the type of loss. A supervised regression problem could be explored in a future work to predict an estimated loss value.} This study introduces a promising approach for predicting propagation loss in hollow-core fibers using supervised machine learning, with the potential to lay the foundation for designing the next generation hollow-core anti-resonant fibers.

\section*{Acknowledgment}
The authors would like to thank Dr. J. E. Antonio-Lopez, Matthew Cooper, Dr. Gregory Jasion, and Dr. Eric Numkan Fokoua.

\bibliography{References}

\bibliographystyle{ieeetr}

\begin{IEEEbiography}[{\includegraphics[width=1in,height=1.25in]{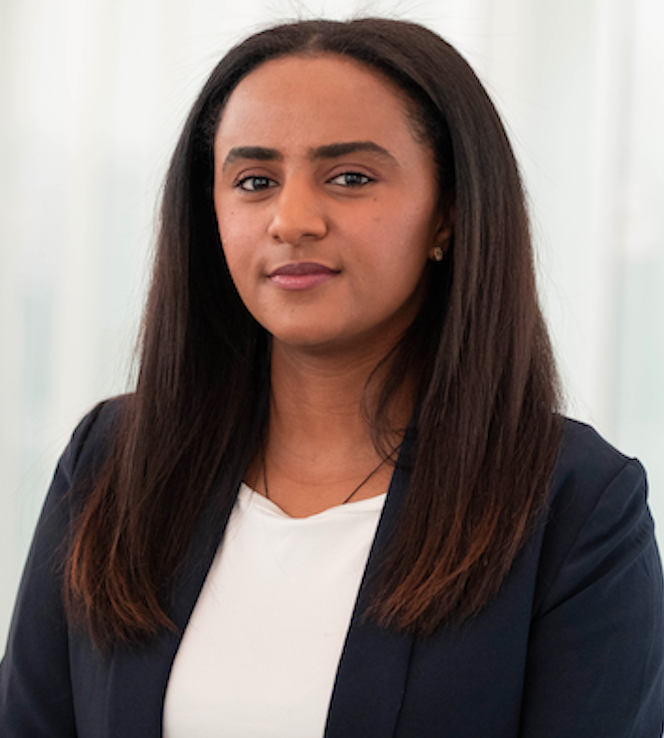}}]
{Yordanos Jewani} earned a Bachelor of Science degree in Electrical and Computer Engineering from Haramaya Institute of Technology in Ethiopia. Currently, she is enrolled in the Data Science Track of the Master's program in Computer Science at Florida Polytechnic University (FPU) and is expected to graduate in May 2023. She is an active member and the student chapter president of the Industrial Engineering and Operation Management (IEOM) Association. From 2018 to 2021, she worked as a database administrator at Catholic Relief Service. Currently, she is working as a graduate assistant in the Department of Data Science and Business Analytics at Florida Polytechnic University. Her research interests include machine learning, computational analysis, optimization, and big data analysis.

\end{IEEEbiography}

\begin{IEEEbiography}[{\includegraphics[width=1in,height=1.25in]{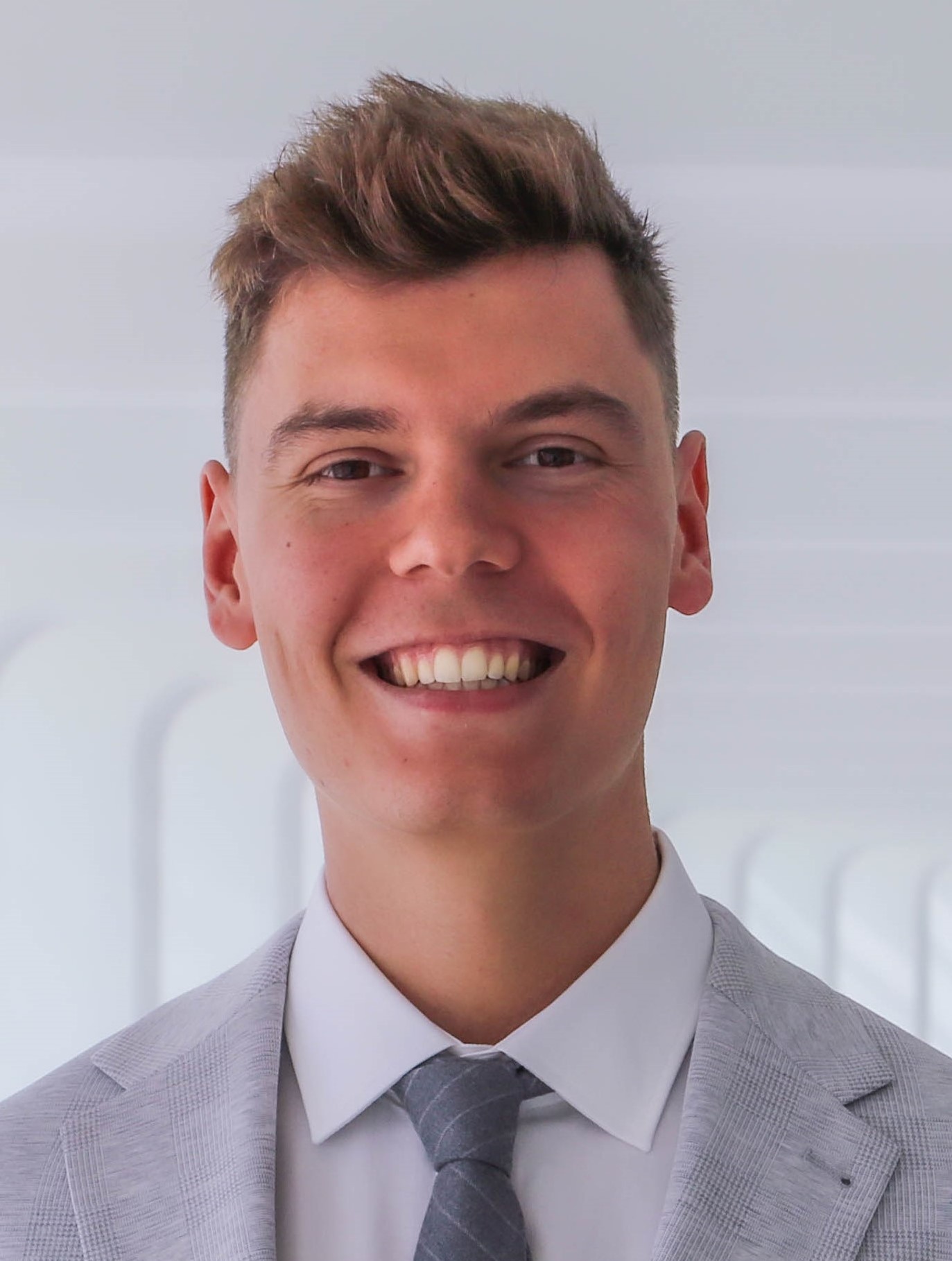}}]{Michael Petry} (S’21) received his B.Eng. degree (with Distinction) in Electrical Engineering from the Karlsruhe University of Applied Sciences (HKA), Germany in 2021. He is currently pursuing a double-master's program in Electrical Engineering at Florida Polytechnic University (FPU) and HKA as a scholar of the German-American Fulbright exchange program. Awarded with the M.Sc. degree at FPU in May 2022, his final graduation is expected in February 2023. As a member of the university's optical fiber group, his research focuses on single- and multi-mode fiber design, statistical characterization and optimization. He is also working on incorporating Machine Learning (ML) models into the HC-ARF domain to provide an alternative approach for fiber analysis. Further fields of interest are information theory and coding using Deep Learning (DL), Radio Frequency (RF), Finite-Element (FE) techniques, VLSI Design, and secure computer architecture. He is an IEEE Photonics Society Member and an IEEE Graduate Student Member.
\end{IEEEbiography}

\begin{IEEEbiography}[{\includegraphics[width=1in,height=1.25in]{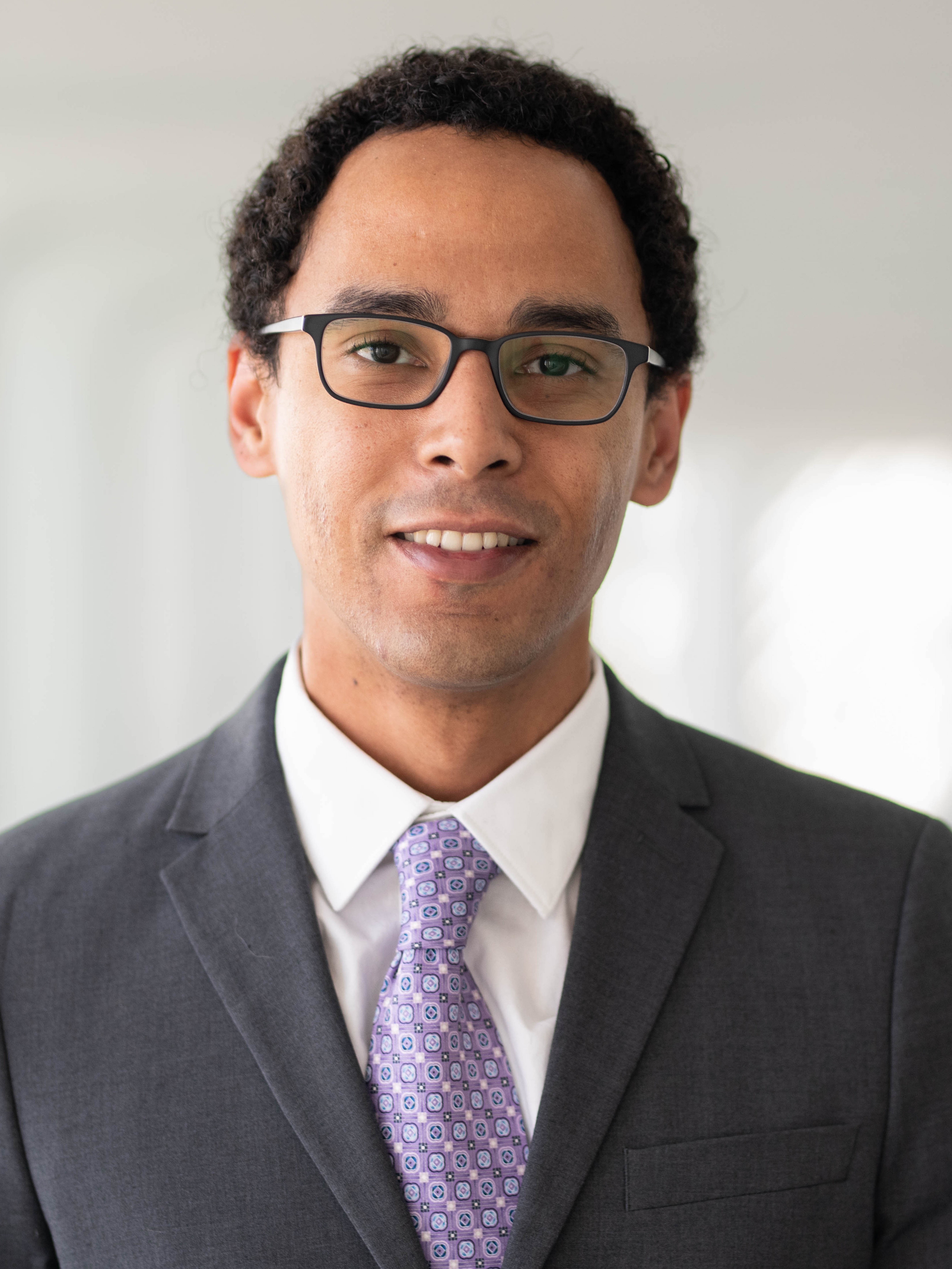}}]{Reinaldo (Rei) Sanchez-Arias}  earned his Bachelor of Science degree in Mathematics from Universidad del Valle in Cali, Colombia. In Fall 2008 he started his doctoral studies in the Computational Science Program at The University of Texas at El Paso (UTEP). During his years at UTEP he was involved in research projects for the Army High Performance Computing Research Center (AHPCRC) in collaboration with a group at Stanford University. He obtained a Ph.D. degree in Computational Science from UTEP in the spring of 2013, working on sparse representation methods for classification problems and dimensionality reduction. He completed a postdoctoral researcher appointment for the AHPCRC working in reduced order models for underbody-blast simulations and data compression techniques. Since August 2018, he is part of the Department of Data Science and Business Analytics at Florida Polytechnic University in Lakeland, FL. Dr. Sanchez-Arias was the recipient of Florida Poly’s Excellence in Teaching Ablaze Award in 2020. He also currently serves as the Assistant Chair of the Department of Data Science and Business Analytics. His general areas of interest include data mining and machine learning, computational linear algebra and optimization, and data science education. His work has been presented at international and national conference meetings including SIAM annual meetings, the International Conference for High Performance Computing, the IEEE International Conference of the Engineering in Medicine and Biology Society, the IEEE International Conference in Machine Learning and Applications, and INFORMS annual meeting. 
\end{IEEEbiography}

\begin{IEEEbiography}[{\includegraphics[width=1in,height=1.25in]{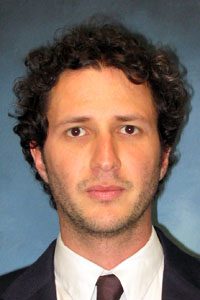}}]{Rodrigo Amezcua-Correa} (Member, IEEE)  received the Ph.D. degree from Optoelectronics Research Centre (ORC) from the University of Southampton, Southampton, U.K. in 2009. Since 2011, he has been with the College of Optics and Photonics, University of Central Florida, Orlando, FL, USA, where he
currently is a Professor of optics and photonics. His research interests include fiber design and fabrication for applications including communications, fiber lasers, nonlinear optics, and sensing.
\end{IEEEbiography}

\begin{IEEEbiography}[{\includegraphics[width=1in,height=1.25in]{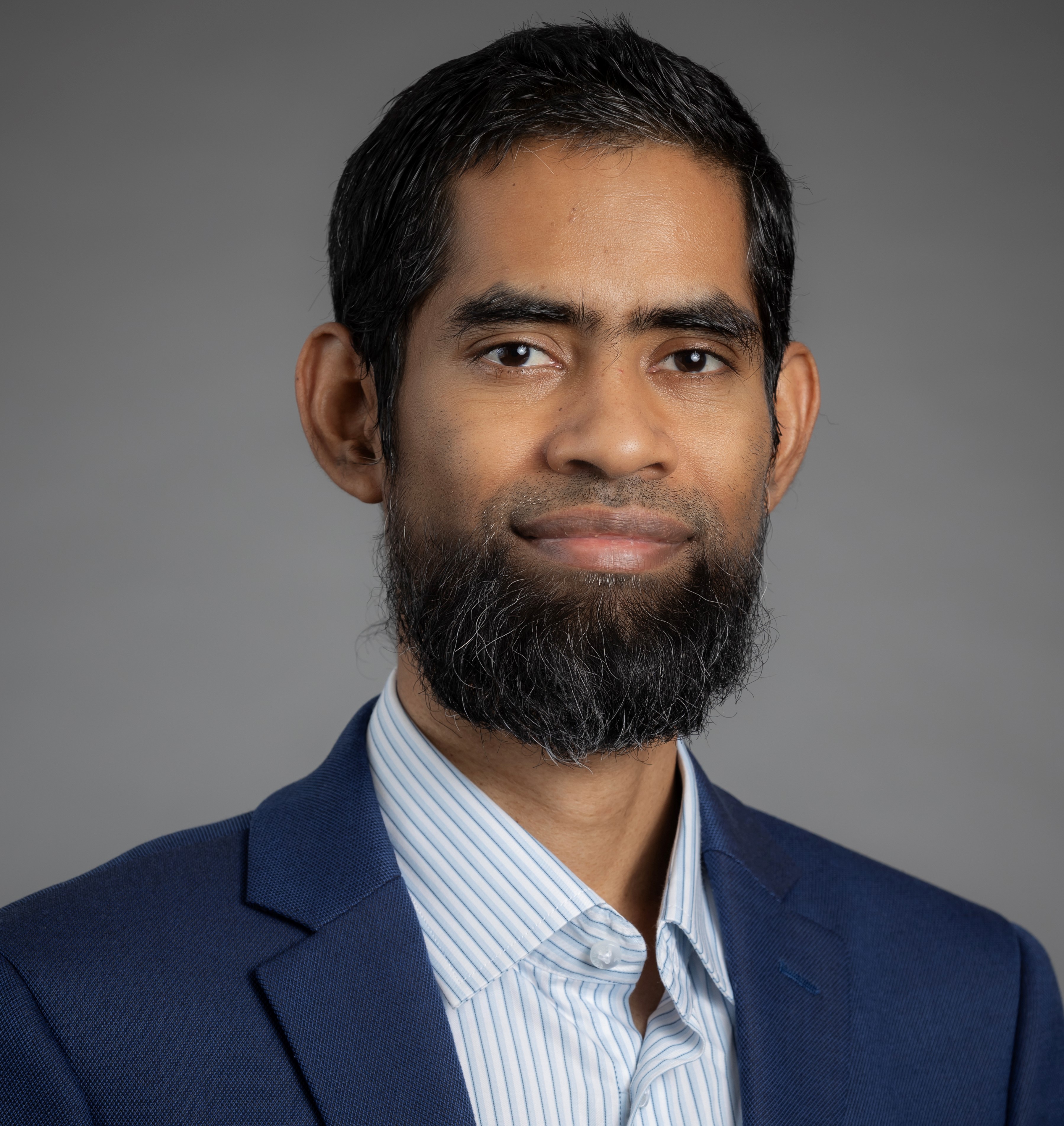}}]{Md Selim Habib} (S'13--SM'19)  received the Ph.D. degree in Electrical and Photonics engineering from the Technical University of Denmark (DTU) in 2017. Following the completion of his doctoral studies, he joined the Fibers Sensors and Supercontinuum Group at the Department of Electrical and Photonics Engineering, DTU, as a Postdoctoral Researcher. After concluding his Postdoctoral Fellowship at DTU, he served as a Postdoctoral Research Associate at CREOL, The College of Optics and Photonics, University of Central Florida, USA, from 2017 to 2019. From 2019 to 2023, Dr. Habib held the position of Assistant Professor of Electrical and Computer Engineering at Florida Polytechnic University, USA. Currently, he is an Assistant Professor of Electrical Engineering at Florida Institute of Technology. His research mainly focuses on computational electromagnetics, emerging optical fiber design, fabrication, and characterization, and ultrafast nonlinear optics. He has published more than 50 articles in refereed journals with over 2250 citations and an h-index of 26. 

Dr. Habib is a Senior Member of Institute of Electrical and Electronics Engineers (IEEE) and Optica (formerly OSA), and Executive officer of OSA Fiber modeling and Fabrication group. Dr. Habib is an Associate Editor of IEEE Access, and Feature Editor of Applied Optics (OSA). He received the University Gold Medal Award from Rajshahi University of Engineering and Technology in 2014.
\end{IEEEbiography}



\end{document}